\documentclass[aps,pra,amsmath,amssymb,floatfix,twocolumn,superscriptaddress]{revtex4}
\usepackage{graphicx}

\newcommand{\ket}[1]{|#1\rangle}
\newcommand{\exv}[3]{\left\langle{#1}\right\vert{#2}\left\vert{#3}\right\rangle}
\newcommand{\ro}{{\bf r}}
\newcommand{\rp}{{\bf r'}}
\newcommand{\Ra}{{\bf R}_A}
\newcommand{\Rb}{{\bf R}_B}
\newcommand{\dint}{\int\!\!\int}
\newcommand{\drdr}{\,d{\bf r} d{\bf r'}}
\newcommand{\dr}{\,d{\bf r}}
\newcommand{\abs}[1]{\left|#1\right|}
\newcommand{\hal}{\frac{1}{2}}

\begin{document}

\title{
Quasiparticle energies for large molecules: a tight-binding GW approach
}
\author{T.A. Niehaus}
\affiliation{ Dept. of Theoretical Physics, 
          University of Paderborn, 
          D - 33098 Paderborn, Germany }
\author{M. Rohlfing}
\affiliation{International university of Bremen, School of Engin. \&
  Science,P.O. Box 750561,D-28725 Bremen, Germany }
\author{F. Della Sala}
\affiliation{ National Nanotechnology Laboratories of INFM, 
           Universit\'a di Lecce, 
           Distretto Tecnologico, 
           Via Arnesano, 73100 Lecce, Italy}
\author{A. Di Carlo}
\affiliation{ INFM and Dept. of Electronic Engineering,
          University of Rome ``Tor Vergata'', 
          00133 Rome, Italy}
\author{  Th. Frauenheim }
\affiliation{ Dept. of Theoretical Physics, 
          University of Paderborn, 
          D - 33098 Paderborn, Germany }
\date{\today}
\begin{abstract}
We present a tight-binding based GW approach for the calculation of
quasiparticle energy levels in confined systems such as molecules. Key
quantities in the GW formalism like the microscopic dielectric function or the
screened Coulomb interaction are expressed in a minimal basis of spherically
averaged atomic orbitals. All necessary integrals are either precalculated or
approximated without resorting to empirical data. The method is validated
against first principles results for benzene and anthracene, where good
agreement is found for levels close to the frontier orbitals. Further, the
size dependence of the quasiparticle gap is studied for conformers of the
polyacenes ($C_{4n+2}H_{2n+4}$) up to $n=30$.
\end{abstract}

\maketitle

\section{Introduction}
Density functional theory (DFT) has nowadays become the standard tool for the
description of ground state properties of such different systems as atoms,
molecules, clusters or bulk materials. Part of the success stems from the fact
that in the DFT electron correlation is in principle exactly covered at the
level of an effective one-particle Hamiltonian. Difficulties arise however,
when the orbital energies obtained from the solution of the Kohn-Sham (KS)
eigenvalue problem are interpreted as approximate quasiparticle energies,
i.e., the energies associated with addition or removal of an electron. A well
known example is the severe underestimation of the band gap in insulating
solids or semiconductor crystals \cite{bec92I}. The same problem is also
present in molecules.  Although it has been shown that Koopmans theorem also
holds for the highest occupied molecular orbital (HOMO) in DFT
 \cite{alm85}, ionization potentials come out too small in practical
calculations. This fact has been traced back to the wrong asymptotic behavior
of common approximate exchange-correlation (XC) functionals like the local
density approximation (LDA)
 \cite{lee94}.  However, the KS gap can be shown to represent a first-order
approximation to optical excitation energies \cite{goe96}, and thus the KS gap
will be different from the quasiparticle gap even if the exact XC functional
is used.

As an alternative to DFT, many body perturbation theory in the approximation
of Hedin \cite{hed65} has been extremely successful in the prediction of
quasiparticle spectra. In this so-called GW method the description of the
electron-electron interaction is approached in a different way than in DFT. While
the exchange energy is exactly calculated like in Hartree-Fock theory, correlation is accounted for by an
energy dependent dielectric function which reduces the coulomb
interaction between electrons. The scheme is nearly self-interaction
free and provides asymptotically correct 
potentials. Consequently, the band gap problem of the DFT is absent in the GW
and also ionization potentials and electron affinities of molecules are
computed in good accord with experiment \cite{roh98,roh00,oht01,ish04}.

In conjunction with the solution of the Bethe-Salpeter (BS) equation
\cite{noz64,sha66,str84}, the GW approximation can also be used to calculate
charge neutral quasiparticle excitations , i.e., optical spectra. In this
context,  the
time-dependent generalization of DFT \cite{gro90}, which in principle
provides the same information as GW+BS has in practice been found to give
inferior results for the absorption spectrum of solids \cite{oni02}. Also for
molecules, time dependent DFT fails in the description of charge transfer excitations
\cite{dre03}, which was attributed to the locality of common XC-functionals
\cite{wan04,taw04,gri04}. Again, GW+BS which contains the correct long range
electron-hole interaction should be able to remedy the problem.

Another example where DFT orbital energies are widely used but lead to
problematic results is given by transport calculations in molecular
devices. Here the calculated currents differ typically by
orders of magnitudes from the experimentally found values
\cite{ree97,div00,pec04}. Since the current-voltage characteristics of such systems
depend critically on the HOMO-LUMO gap, a major improvement is expected when
GW quasiparticle energies would be used instead of DFT orbital energies.

It is clear from the forgoing discussion, that although the GW approximation
was developed in the context of solid state theory, its application to
molecular systems is becoming more and more important. In fact,
implementations for systems with translational symmetry using plane wave basis
sets can directly be applied to finite systems when very large super-cells are
used. Nonetheless, the use of localized atomic-like basis sets is clearly more
adapted to the problem and such implementations have been utilized quite
successful in the past years \cite{roh95,roh96,roh98,roh00}. But even with
this improvement, the numerical complexity of the GW equations limits a first
principles evaluation to rather small system sizes of tens of
atoms. So applications like 
transport calculations for molecules, where a sizable amount of the atoms
comprising the leads need to be included, or molecular dynamics in the excited
state are currently not feasible. It
would therefore be desirable to have an approximate GW scheme which
nevertheless captures the essential physics of the underlying theory. The
purpose of this paper is to propose such a scheme.

In the past years a number of different approximated GW methods have been
introduced and successfully applied
\cite{hed65,hyb86,bec92,pal95,ste84,bec88,gu94,del97,del00,fur02}. The main
difference to this earlier work is that we focus on molecular applications
with a real space implementation. Furthermore we try to avoid empirical
parameters in order to achieve a higher transferability.
 
Like GW calculations usually employ DFT energies and wavefunctions as zero
order approximations to the quasiparticle quantities, our approach is based on
the {\em Density Functional based Tight-Binding} (DFTB) method
\cite{por95,els98}, which itself is an approximation to DFT. The DFTB scheme
has been shown to provide reliable results on a variety of systems classes
ranging from molecules to solids at a highly reduced numerical cost compared
to DFT calculations. In section \ref{meth} the DFTB method is briefly
introduced together with a detailed description of the approximations in the
various quantities involved in the GW formalism. The accuracy and main
shortcomings of our approach are then examined in section \ref{appl}, where it
is applied to a prototype series of $\pi$-bonding molecules, the polyacenes.

\section{Methodology}
\label{meth}
The GW method has been extensively discussed in the literature and several
reviews are available \cite{ary98,aul99,far99,hed99,oni02}. The main goal is
to solve the Dyson equation:
\begin{equation}
  \label{dyson}
 \left( H_0 + \Sigma(\epsilon^{QP}_i)\right) \ket{\psi_i^{QP}} =
 \epsilon^{QP}_i \ket{\psi_i^{QP}}, 
\end{equation}
for the quasiparticle energies $\epsilon^{QP}_i$ and wavefunctions
$\ket{\psi_i^{QP}}$. Here $H_0$ is the Hartree Hamiltonian and the so-called
self-energy $\Sigma$ is a nonlocal and energy dependent operator, which
accounts for exchange and correlation effects. It thus can be seen as a
replacement for the local exchange-correlation potential $v_{xc}$ in the DFT
Kohn-Sham (KS) equations. In the GW approximation of Hedin, the
self-energy is given as a product of the single-particle Greens function $G$ and
the screened Coulomb interaction $W$. As these quantities, as well as the
Hartree Hamiltonian, depend on the quasiparticle wavefunctions, the Dyson
equation [Eq.~(\ref{dyson})] has to be iterated until self-consistency is
achieved. However, since the DFT one-particle wavefunctions are usually very
similar to the final quasiparticle ones and moreover self-consistency not
necessarily improves the result \cite{sch98}, Eq.~(\ref{dyson}) may be simplified to
\begin{equation}
  \label{qpe}
  \epsilon_i^{QP} = \epsilon_i^{DFT} +
      Z_i\langle \psi_i|\Sigma(\epsilon_i^{DFT})-v_{xc}|\psi_i\rangle,
\end{equation}
where $\psi_i$ are KS orbitals, non-diagonal elements of the Dyson Hamiltonian in the basis of DFT
states are neglected and the energy dependence of the self-energy has
partially been accounted for by the renormalization constant $Z_i$:
\begin{equation}
  \label{reno}
 Z_i = \left(1-  \left.\frac{\partial \Sigma(\omega)}{ \partial
 \omega}\right|_{\epsilon_i^{DFT}}\right)^{-1}.
\end{equation}

In our approach Eq.~(\ref{qpe}) is now subjected to further approximations which
are presented separately for each term in the following.

\subsection{The DFT orbital energies $\epsilon_i^{DFT}$}
\label{DFTBeps}
As already mentioned, the Kohn-Sham energies $\epsilon_i^{DFT}$ and orbitals are
obtained from the DFTB method, which has been presented in detail earlier (for
a review see Ref.~\cite{fra02}). Here we only describe the method to the
extent necessary to motivate the remaining approximations in this work. In the
DFTB, the Kohn-Sham states $\psi_i^{\text{DFTB}}$ are expanded in a linear
combination of atom centered orbitals $\phi_{\mu}$:
\begin{equation}
  \label{DFTBbase}
   \psi_i(\ro) = \sum_{\mu}c_{\mu i}\phi_{\mu}(\ro-\Ra),
\end{equation}   
which are obtained from a preceding DFT calculation on neutral atoms. Here
$\mu:=\{Alm\}$ is a compound index indicating the atom on which the
basisfunction is centered, its angular momentum $l$ and magnetic quantum
number $m$. Later, also quantities which depend only on $A$ and $l$
appear. These will be denoted with a corresponding index $\bar{\mu}:=\{Al\}$
throughout the paper.

Since atomic orbitals are usually too long ranged to be used directly in a molecular
calculation, the atomic DFT Hamiltonian is augmented with a confining square
potential to compress the wavefunction outside of a given radius $r_0$ (usually
twice the covalent radius of the element), while ensuring the desired cusp
conditions inside \cite{por95}. From the atomic valence states a minimal basis of $s$ and
$p$ orbitals is then chosen, although also $d$-orbitals are included when
necessary, e.g.~for second row elements \cite{nie01}. With the help of the expansion
(\ref{DFTBbase}) the Kohn-Sham equations of DFT can be written:
\begin{subequations}
 \begin{gather}
  \label{kse}
    \sum_\nu c_{\nu i} (H_{\mu\nu} - \epsilon_i S_{\mu\nu}) =
  0 , \;\;\; \forall\;\mu ,\:i\;\\
 S_{\mu\nu} = \langle\phi_\mu|\phi_\nu\rangle\label{smunu} \\
 H_{\mu\nu} = H_{\mu\nu}^0+H_{\mu\nu}^{SCC},\label{hmunu}
\end{gather}  
\end{subequations}
where the overlap matrix $S_{\mu\nu}$ has been introduced and the Hamiltonian
is divided in two parts. The first part $H_{\mu\nu}^0$ is approximated as
follows:
 \begin{eqnarray}
\label{DFTBme}
H_{\mu \nu}^0 = 
\left\{ 
\begin{array}{c@{\quad :\quad}l}
 \epsilon_{\mu}^{\text{free atom}} & \mu = \nu \\
 \langle\phi_{\mu} ( { \ro }) |  H_{DFT}[\rho_{A}^0+\rho_{B}^0] |\phi_{\nu} (\ro)
  \rangle  &\mu\, \epsilon\, A, \nu\, \epsilon\, B \\
 0 & \text{otherwise}
\end{array}
\right.
\end{eqnarray} 

The KS Hamiltonian in Eq.~(\ref{DFTBme}) contains as usual the kinetic
energy, the electron-nuclei attraction and the  Hartree as well as
exchange-correlation potential and depends only on the atomic densities of
atoms $A$ and $B$. This means, that besides the crystal field terms also
all three-center terms are neglected. The onsite elements are given as atomic
orbital energies obtained from a DFT calculation without confining potential
to ensure the right dissociation limit.  The integrals in Eq.~(\ref{DFTBme})
are numerically evaluated and tabulated for varying distance between atoms $A$
and $B$. 

The second part of the Hamiltonian (\ref{hmunu}) corrects for the fact that the
molecular density differs from a simple superposition of atomic
densities. In order to estimate this difference, spherical averages over
basis functions belonging to one angular momentum shell are build
\begin{equation}
  \label{sbas}
  F_{Al}(\ro) = \frac{1}{2l+1} \sum_{m=-l}^{m=l} \left| \phi_{Alm}(\ro) \right|^2,
\end{equation}
and used in a Mulliken type approximation
\begin{equation}
  \label{mul}
  \phi_\mu(\ro) \phi_\nu(\ro) \approx \hal S_{\mu\nu} \left(
  F_{\bar{\mu}}(\ro) +   F_{\bar{\nu}}(\ro) \right), 
\end{equation}
to represent the molecular density. The latter is then constructed from point
charges
\begin{gather}
  \label{rhoq}
  \rho(\ro) = \sum_i\abs{\psi_i(\ro)}^2 \approx \sum_{\bar{\mu}} q_{\bar{\mu}}
  F_{\bar{\mu}}(\ro) \nonumber \\ \text{with\quad}
   q_{\bar{\mu}} =
 \sum_{m=-l}^{l}\sum_{\nu i} 
   c_{\mu i}
         c_{\nu i} 
       S_{\mu\nu};
 \end{gather}
an expansion, which despite of its simplicity takes the different spatial
localization of e.g $s$ and $p$-orbitals into account. Based on these
considerations, the difference between the true molecular density and
superimposed atomic densities can be estimated with net Mulliken charges
$\Delta q_{\bar{\mu}} = q_{\bar{\mu}} - q_{\bar{\mu}}^{\text{atom}}$ and leads
to the correction term \cite{rem1}:
\begin{equation}
H_{\mu\nu}^{SCC}= \hal
  S_{\mu\nu}\sum_{\bar{\delta}}(\gamma_{\bar{\mu}\bar{\delta}}+\gamma_{\bar{\nu}\bar{\delta}})\Delta
  q_{\bar{\delta}},
\end{equation}
as shown in more detail in Ref.~\cite{els98}.

The term $\gamma$ describes the interaction of two electrons in the orbitals
$\bar{\mu}$ on atom $A$ and $\bar{\nu}$ on atom $B$, including the effects of
exchange and correlation:
\begin{subequations}
 \begin{eqnarray}
  \label{gamma}
  \gamma_{\bar{\mu}\bar{\nu}} &=&  \dint F_{\bar{\mu}}(\ro)\left( \frac{1}{|\ro-\rp|}
 +\frac{\delta v_{xc}[{\rho(\ro)}]}{\delta \rho(\rp)} \right)
  F_{\bar{\nu}}(\rp) \drdr\nonumber\\\\
         &\approx&  \gamma_{\bar{\mu}\bar{\nu}}(\abs{\Ra-\Rb},U^H_{\bar{\mu}},U^H_{\bar{\nu}}),
\end{eqnarray} 
\end{subequations}
and is approximated by considering two limiting cases. For large distances
between the two atoms, the integral (\ref{gamma}) simplifies to a pure Coulomb
interaction of two point charges, since the $v_{xc}$ contribution dies off
rapidly. For short distance on the other hand, Eq.~(\ref{gamma}) becomes an
atomic integral $U_H$, which can be easily calculated numerically for each
element.  From these limiting cases a simple interpolation formula was derived
in Ref.~\cite{els98}, which is a function of the atomic parameters $U^H$
and the interatomic distance only. Since the Mulliken net charges $\Delta
q_{\bar{\delta}}$ depend on the molecular orbital coefficients $c_{\mu i}$,
Eq.~(\ref{kse}) has to be iterated until self-consistency. As a result the
orbital energies $\epsilon_i^{\text{DFTB}}$ needed in Eq.~(\ref{qpe}) are
obtained.
\subsection{The self-energy $\Sigma_i(\epsilon)$}
\label{self}
We calculate the self-energy in the GW approximation by:
\begin{equation}
  \label{seli}
  \Sigma(\ro,\rp,\epsilon) = \frac{i}{2\pi}\int  e^{i\omega 0^+}
  G_0(\ro,\rp,\epsilon-\omega)\; W(\ro,\rp,\omega)\;d\omega,
\end{equation}
where $G_0$ is the single particle Greens function built from DFTB
wavefunctions and $W=\epsilon^{-1}v$ is the screened Coulomb interaction,
while $v$ is the bare one. For
the following it is beneficial to divide the self-energy into two parts as
$\Sigma = iG_0v + iG_0(\epsilon^{-1}-1)v$. The latter term denoted $\Sigma^c$
is energy dependent and describes dynamical correlation effects, while the
former term $\Sigma^x$ provides the major part of the self-energy. For
$\Sigma^x$ the frequency integration in Eq.~(\ref{seli}) can be carried out
easily and yields in the KS basis the usual Hartree-Fock exchange energy for
orbital $i$:
\begin{equation}
  \label{x}
  \Sigma^x_i = \sum_j^{occ} \dint
  \frac{\psi_i(\ro)\psi_j(\ro)\psi_i(\rp)\psi_j(\rp)}{|\ro-\rp|} \drdr.
\end{equation}
Since in contrast to empirical tight-binding schemes the basis functions
$\phi_\mu$ are available in the DFTB method, one could in principle calculate
Eq.~(\ref{x}) directly from the wavefunctions. In this way the method would
scale like first principles schemes with $N^4$, where $N$ is the number of
basis functions.  Therefore we seek for an approximate solution and
note that after expansion of the KS states in atomic orbitals,
Eq.~(\ref{x}) contains products of basis functions which are in
general located
on different atomic centers. An important simplification can thus be achieved,
when the Mulliken
approximation [Eq.~(\ref{mul})] is applied to  the integral. Introducing the
following notation for the matrix elements of a general two-point function in
the basis of squared and spherically averaged DFTB atomic orbitals:
\begin{equation}
  \label{twop}
  [f]_{\bar{\mu}\bar{\nu}} = 
 \dint  F_{\bar{\mu}}(\ro) f(\ro,\rp) F_{\bar{\nu}}(\rp) \drdr,
\end{equation}
we then arrive at the following simplified expression for $  \Sigma^x_i $:
\begin{equation}
  \label{sigx}
   \Sigma^x_i = \sum_j^{occ} \sum_{\bar{\mu}\bar{\nu}} 
q^{ij}_{\bar{\mu}} \left[v\right]_{\bar{\mu}\bar{\nu}} q^{ij}_{\bar{\nu}} . 
\end{equation}
Here the $ q^{ij}_{\bar{\mu}}$ are generalized Mulliken charges
\begin{equation}
  \label{qij}
          q^{ij}_{\bar{\mu}} =
 \frac{1}{2}
\sum_{m=-l}^{l}  \sum_\nu 
  \left( c_{\mu i}
         c_{\nu j} 
        S_{\mu\nu} + 
        c_{\nu i}
        c_{\mu j} S_{\nu\mu} \right),
\end{equation}
which provide a point charge representation of the overlap between two
molecular orbitals $i$ and $j$. The important observation is now that the
matrix of the Coulomb interaction is equal to the definition of the
$\gamma$-functional in Eq.~(\ref{gamma}), when the contributions stemming from
the XC functional are removed. In other words, the functional form of
$\gamma$ can also be used for $[v]$, if the atomic parameters $U^H$ are
replaced by the parameters $U^{ee}$ which incorporate only the classical
Coulomb interaction. We calculate these electron repulsion integrals directly
from the DFTB basis functions using the algorithms presented in
Ref.~\cite{gus02}. The parameters for each angular momentum are set to
an average over the integrals for different combinations of the magnetic
quantum numbers belonging to that shell.

The main drawback of the Mulliken approximation in Eq.~(\ref{mul}) is, that
onsite integrals of the exchange type are completely neglected. These,
however, contribute around 10 \% to the final exchange energy. Similar to the
proceeding in the quantum chemical INDO approach \cite{rid73} we therefore
include all non-vanishing onsite integrals, leading to the following final
form for $  \Sigma^x_i $ \cite{rem2}:
\begin{gather}
  \label{sigxf}
   \Sigma^x_i = \sum_j^{occ} \sum_{\bar{\mu}\bar{\nu}} q^{ij}_{\bar{\mu}}
   \left[v\right]_{\bar{\mu}\bar{\nu}} q^{ij}_{\bar{\nu}} \nonumber\\+ \sum_A
   \sum_{\mu,\nu \ni A}^{\mu\ne\nu} \left(
   c_{\mu i}^2  c_{\nu j}^2 +  c_{\mu i} c_{\nu j} c_{\nu i} c_{\mu j} \right)
   (\phi_\mu \phi_\nu|\phi_\mu \phi_\nu). 
\end{gather}
While in the INDO approach the necessary integrals are taken as empirical
fitting parameters, we compute them from the atomic basis functions. More
precisely, the parameters are calculated from an uncompressed wavefunction in
order to be consistent with the onsite definition of the DFTB Hamiltonian
matrix elements.  The values used in this study are given in
Tab.~\ref{tab_par} together with the $U^H$ and $U^{ee}$ parameters.
\begin{table}
\caption{The atomic electron-electron interaction integrals $U^H_l$ and
  $U^{ee}_l$, as well as the exchange integrals $(\phi_{lm}
  \phi_{l'm'}|\phi_{lm} \phi_{l'm'})$ used in this study. Results are given for
  free and compressed atomic basisfunctions, as defined by the confinement
  radius $r_0$ (see Sec.~\ref{DFTBeps} and Ref.~\cite{por95}). The
  same compression is used in the calculation of the Hamiltonian and overlap
  matrix elements. \label{tab_par}}
\begin{ruledtabular}
\begin{tabular}{lccc}
Element & Parameter & $r_0$ [$a.u.$] & Value [eV]\\\colrule
Hydrogen&&&\\
&  $U^H_0$ & $\infty$ & 11.06 \\
&  $U^{ee}_0$ & $\infty$ & 15.39 \\ 
&  $U^{ee}_0$ & 3.0 & 21.36 \\ 
Carbon&&&\\
&  $U^H_0$ & $\infty$ & 10.81 \\
&  $U^H_1$ & $\infty$ & 10.81 \\
&  $U^{ee}_0$ &$\infty$ & 15.66 \\
 &  $U^{ee}_1$ & $\infty$ & 14.15 \\
&  $U^{ee}_0$ & 2.7 & 17.98 \\
 &  $U^{ee}_1$ & 2.7 & 18.72 \\
& $(\phi_{00} \phi_{1m'}|\phi_{00} \phi_{1m'})$ &  $\infty$ &3.01 \\
& $(\phi_{1m} \phi_{1m'}|\phi_{1m} \phi_{1m'})$ &  $\infty$ &0.75 
\end{tabular}
\end{ruledtabular}
\end{table}

Let us now turn to the correlation contribution of the self-energy $\Sigma^c$,
which is much harder to evaluate, since it amounts to a multi step
procedure. First we construct matrix elements of the electronic polarizability
in the random-phase approximation according to:
\begin{gather}
  \label{pol}
  [P(\omega)]_{\bar{\mu}\bar{\nu}} = 2 \sum_k^{occ} \sum_l^{virt}
  \left(\sum_{\bar{\alpha}} \tilde{S}_{\bar{\mu}\bar{\alpha}}
  q^{kl}_{\bar{\alpha}} \right) \left(\sum_{\bar{\beta}} q^{kl}_{\bar{\beta}}
  \tilde{S}_{\bar{\beta}\bar{\nu}} \right) \times \\ \left[
  \frac{1}{\epsilon^{\text{\tiny DFTB}}_k-\epsilon^{\text{\tiny DFTB}}_l
  -\omega +i0^+} + \frac{1}{\epsilon^{\text{\tiny
  DFTB}}_k-\epsilon^{\text{\tiny DFTB}}_l + \omega +i0^+} \right],\nonumber
\end{gather}
where we again used the Mulliken approximation of Eq.~(\ref{mul}) and
introduced the overlap matrix of spherically averaged DFTB basis functions $
\tilde{S}_{\bar{\mu}\bar{\nu}}=\int  F_{\bar{\mu}}(\ro) F_{\bar{\nu}}(\ro) d\ro$,
not to be confused with the overlap appearing in the KS equations
(\ref{smunu}). 

The quantity $\tilde{S}$ never needs to be constructed, since it appears only
in intermediate quantities and falls out in the final equation for the
screened Coulomb interaction we are aiming at.

In a next step we obtain the dielectric function in matrix notation as:
\begin{equation}
  \label{eps}
  [\epsilon(\omega)] = \tilde{S} - [v]\, \tilde{S}^{-1}\, [P(\omega)].
\end{equation}
For systems with translational symmetry the dielectric matrix is hermitian in
reciprocal space. This fact is used in plasmon-pole models
\cite{hyb86,lin88,god89} to simplify the frequency integration in
Eq.~(\ref{seli}), which is numerically demanding due to the complicated pole
structure of $\epsilon^{-1}$ along the real axis. In these models the inverse
of the dielectric matrix is diagonalized and the eigenvalues are assumed to be
a simple function of the frequency, while the eigenvectors are frequency
independent. Free parameters of the model are either obtained from sum rules
or by diagonalizing $\epsilon^{-1}$ at different test frequencies. It is then
easy to perform the frequency integration analytically to obtain the
self-energy.

However, in the present real-space approach the inverse dielectric matrix is
not symmetric. In Ref.~\cite{roh95} this problem was circumvented by
introducing an auxiliary symmetrized dielectric matrix, while we proceed by
noting that the screened Coulomb interaction $W$:
\begin{equation}
  \label{W}
  [W(\omega)] =  \tilde{S}\,[\epsilon(\omega)]^{-1}\,[v],
\end{equation}
has the desired property of being symmetric. Applying the plasmon-pole
approximation to $[W-v]$, we finally arrive at the following expression for the
correlation contribution to the self-energy for orbital $i$:
\begin{multline}
  \label{wmv}
  \Sigma^c_i(\omega)= \sum_n\sum_{\bar{\delta}} \left(
 \sum_{\bar{\mu}} q^{in}_{\bar{\mu}} \Phi_{\bar{\mu}\bar{\delta}} \right)^2 \times \\
\frac{z_{\bar{\delta}}
 \omega_{\bar{\delta}}}{2}
\left\{ 
\begin{array}{c@{\quad :\quad}l}
 \frac{1}{\omega-\epsilon_n^{\text{DFTB}}+\omega_{\bar{\delta}}} & n \in \text{occ} \\
  \\
 \frac{1}{\omega-\epsilon_n^{\text{DFTB}}-\omega_{\bar{\delta}}}  &  n \in \text{virt},
\end{array}
\right.
\end{multline}
where $\Phi$ denotes the eigenvectors of $[W-v]$ , while $z_{\bar{\delta}}$ and
$\omega_{\bar{\delta}}$ are the mentioned parameters of the plasmon-pole model, as
defined in Ref.~\cite{roh95}. They
are determined by diagonalization of $[W-v]$ at zero frequency and one
frequency on the imaginary axis. We checked that the actual values chosen have
little impact ($< 0.1 $ eV) on the final quasiparticle energies.

Based on the self-energy, the renormalization constant $Z_i$ from
Eq.~(\ref{qpe}) is then obtained by a simple numerical differentiation. For
the molecular structures we studied, $Z_i$ is usually roughly $0.85$, which is
close to the values reported for bulk systems \cite{hyb86}. However, for certain unbound
virtual orbitals, $Z_i$ can decrease to as much as 0.5.

\subsection{The exchange-correlation contribution $v^{xc}_i$} 
\label{vxc}
We complete the description of our method with an investigation of the
contributions to the quasiparticle energies arising from the
exchange-correlation potential, denoted $v^{xc}_i[\rho_v]$. As indicated,
$v_{xc}$ is evaluated at the valence density $\rho_v$, consistent with the
fact that the summation in the exchange energy [Eq.~(\ref{sigx})] is carried out
over valence orbitals $j$ only. As pointed out in Ref.~\cite{ish01}, the
core contribution to the exchange energy is not neglible and this holds also
for the core contribution of $v^{xc}$. However, even for
exchange-correlation potentials commonly used today, which are far from exact,
both core contributions cancel to a large degree when computing quasiparticle
energies according to Eq.~(\ref{qpe}).

In analogy to the derivation of the DFTB method, we now expand  $v^{xc}$
around the density $\rho_v^0$, which is a superposition of atomic valence
densities. With $\rho_v = \rho_v^0+ \delta\rho$ we obtain:
\begin{subequations}
\begin{gather}
  v^{xc}_i[\rho_v] = \int \abs{\psi_i(\ro)}^2 v^{xc}_i[\rho_v^0(\ro)] \,\dr
  \,\,+ \nonumber\\ \dint \abs{\psi_i(\ro)}^2 \,
 \frac{\delta v^{xc}_i[\rho_v(\ro)]}{\delta\rho_v(\rp)} \delta\rho(\rp) 
  \drdr + {\cal O}(\delta\rho^2)\label{vxcia} \\
     \approx  \sum_{\mu\nu} c_{\mu i}c_{\nu i}  v^{xc}_{\mu\nu}[\rho_v^0] +
  \sum_{\bar{\mu}\bar{\nu}} q^{ii}_{\bar{\mu}} 
\left[\frac{\delta v^{xc}}{\delta\rho}\right]_{\bar{\mu}\bar{\nu}} 
 \Delta q_{\bar{\nu}}\label{vxcib}
\end{gather}  
\end{subequations}
In going from Eq.~(\ref{vxcia}) to Eq.~(\ref{vxcib}), the Mulliken
approximation was again employed and matrix elements of the exchange-correlation
kernel $\delta v^{xc}/\delta\rho$ were introduced in the notation of
Eq.~(\ref{twop}). The first term of Eq.~(\ref{vxcib}) is now exactly treated
like the Hamiltonian in the DFTB scheme. That is, only the two-center terms
are kept and the onsite values are calculated from uncompressed basis
functions and atomic densities. Then the integrals are numerically evaluated
and tabulated in the usual Slater-Koster form as a function of interatomic
distance. 

\begin{figure} 
  \centering
 \includegraphics[scale=0.7]{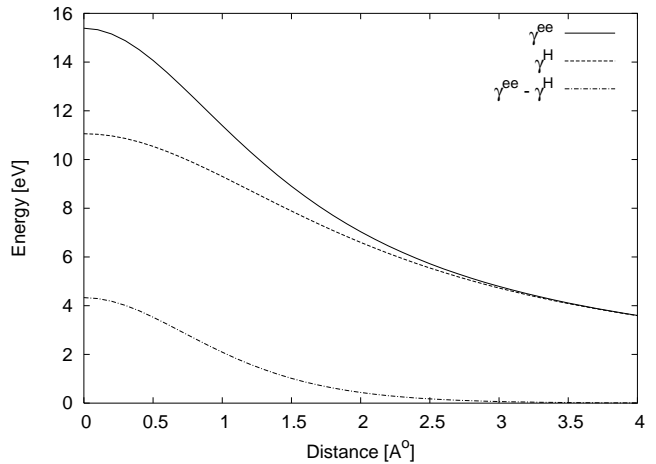}
\caption{The electron-electron interaction integrals for hydrogen as a
 function of distance. Shown are results including both Coulomb and
 exchange-correlation interactions ($\gamma^H$), pure Coulomb
 ($\gamma^{ee}$) and the {\em negative} of the pure exchange-correlation
 interaction ($-\delta v^{xc}/\delta\rho =\gamma^{ee}-\gamma^H$), according to
 Eq.~(\ref{gxc}).\label{gamfig}}
 \end{figure}

The second term in  Eq.~(\ref{vxcib}) is the counterpart of $H_{\mu\nu}^{SCC}$
in Eq.~(\ref{hmunu}). If we set 
\begin{multline}
  \label{gxc}
  \left[\frac{\delta v^{xc}}{\delta\rho}\right]_{\bar{\mu}\bar{\nu}} =
  \gamma_{\bar{\mu}\bar{\nu}}(\abs{\Ra-\Rb},U^H_{\bar{\mu}},U^H_{\bar{\nu}})\, - 
\\ \gamma_{\bar{\mu}\bar{\nu}}(\abs{\Ra-\Rb},U^{ee}_{\bar{\mu}},U^{ee}_{\bar{\nu}}),
\end{multline}
the long range $1/R$ tail of the two $\gamma$-functions cancels (see Fig.~\ref{gamfig}), and one is
left with a short ranged representation of the exchange-correlation kernel
without introducing any new parameters or integral approximations. Moreover
the $v^{xc}$ contribution of the self-energy now cancels all related
contributions in the orbital energies $\epsilon^{\text{DFTB}}_i$ as it should be \cite{rem3}.

At this point, all the necessary ingredients to calculate quasiparticle
energies within the DFTB scheme have been presented. In the next section we
analyze the strengths and weaknesses of the method by applying it to the
polyacene series.
\begin{table*}
\squeezetable
\caption{The different contributions to the quasiparticle (QP) energies and
  the QP energies themselves obtained from the method described in this work (DFTB) as
  well as from first principles calculations using Gaussian type orbitals
  (GTO).  Shown are results for some levels close to the frontier orbitals of
  the benzene and anthracene molecules. All energies in eV.\label{tab_vgl}}
\begin{ruledtabular}
\begin{tabular}{ll|cc|cc|cc|cc|cc}
& &\multicolumn{2}{c}{$\epsilon^i$}& \multicolumn{2}{c}{$v_{xc}^i$} &
\multicolumn{2}{c}{$\Sigma_x^i$} &
\multicolumn{2}{c}{$\Sigma_c^i$}  &
\multicolumn{2}{c}{$\epsilon_{QP}^i$} \\
State & Sym. &  GTO & DFTB  &  GTO & DFTB  &  GTO &
DFTB  &  GTO & DFTB &  GTO & DFTB\\
\colrule
\multicolumn{2}{l|}{Benzene}         \\
$A_{2u}$ & $\pi$     & -9.37& -8.95&-12.82&-12.39&-17.15&-16.35& 2.03& 
2.47&-11.67&-10.45\\
$E_{2g}$ & $\sigma$  & -8.34& -7.71&-15.05&-12.25&-19.49&-15.62& 1.86&1.41 
&-10.92& -9.67\\
$E_{1g}$ & $\pi$     & -6.59& -6.64&-13.03&-12.30&-15.61&-14.98& 0.59&0.87 & 
-8.58& -8.46\\\colrule
$E_{2u}$ &  $\pi^*$  & -1.30& -1.32&-12.64&-11.72& -7.58& -7.41&-1.74&-0.99&  
2.01&  2.00\\
$B_{2g}$ &  $\pi^*$  &  0.92&  2.29& -6.96&-11.19& -2.89& -6.33&-2.31&-2.84& 
2.67&  4.31\\
\colrule\colrule
\multicolumn{2}{l|}{Anthracene}\\
$B_{3u}$ &  $\pi$    & -7.97 & -7.62 &-12.98 &-12.25&-16.32&-15.61& 1.98 &2.27  
&  -9.33 &-8.71\\
$B_{2g}$ &$\sigma$   & -7.85 & -7.28 &-13.07 &-12.17&-16.40&-15.32& 2.01 &1.90   
& -9.15 &-8.53\\
$A_{u }$ & $\pi$     & -6.82 & -6.78 &-13.12 &-12.24&-15.81&-15.11& 1.38 &1.66  
&  -8.12 &-7.98\\
$B_{1g}$ & $\pi$     & -6.51 & -6.40 &-13.30 &-12.10&-15.27&-14.18& 1.01 &1.06  
&  -7.47 &-7.42\\
$B_{2g}$ & $\pi$     & -5.30 & -5.51 &-13.28 &-12.18&-14.90&-14.37& 0.58 &0.88  
&  -6.34 &-6.82\\\colrule
$B_{3u}$ &  $\pi^*$  & -2.86 & -2.97 &-13.08 &-11.86& -8.78& -8.17&-1.82 
&-0.90  & -0.37 &-0.19\\
$A_{u }$ &   $\pi^*$ & -1.58 & -1.59 &-13.18 &-11.62& -8.49& -7.92&-2.21 
&-1.38 &   0.89 & 0.74\\
$B_{1g}$ &  $\pi^*$  & -1.25 & -1.28 &-12.89 &-11.63& -7.63& -7.34&-2.54 
&-1.83 &   1.47 & 1.18\\
$B_{3u}$ &  $\pi^*$  & -0.52 &  0.01 &-11.90 &-11.41& -6.47& -6.94&-2.84 
&-2.45 &   2.07 & 2.03\\
\end{tabular}
\end{ruledtabular}
\end{table*}
\section{Applications}
\label{appl}
\subsection{Comparison to first principles results}
\begin{figure} 
  \centering
 \includegraphics[scale=0.3]{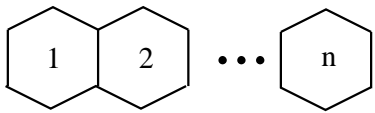}
\caption{Schematic viewgraph of the polyacenes: n is the number of monomers.\label{polyacene}}
 \end{figure}

 The polyacenes ($C_{4n+2}H_{2n+4}$) are linear chains 
 of anellated polycyclic aromatic hydrocarbons, 
 as  shown in Fig. \ref{polyacene}.
 The monomer ($n=1$) is benzene. Naphtalene is with $n=2$, anthracene $n=3$,
 tetracene $n=4$ and so on. These systems have received much attention because
 of their potential use in efficient organic thin film devices. Theoretically
 they have been characterized quite widely
 \cite{wha79,tri91,cio93,sab96,wib97,not98,del03} and recently also an
 investigation in the context of the GW approximation appeared, where
 polymorphs of the pentacene crystal were analyzed in terms of their optical
 spectra \cite{tia03}.

 Here, the polyacenes are chosen as a prototypical $\pi$-system to
 explore the accuracy of our approximations. To this end we fully optimized
 the different structures at the DFTB level of theory without imposing
 symmetry constraints. The obtained geometries are in excellent agreement with
 a recent DFT study on the polyacenes from $n=1$ to $5$, in which the hybrid
 functional B3LYP and the accurate 6-311G** basis set was employed \cite{wib97}. For all
 the molecules studied, we find a mean deviation in the bond lengths of no more
 than 0.005 \AA.
Then the quasiparticle spectrum was obtained according to the approximations
in the last section. For comparison, also first principles GW calculations in
the Gaussian type orbital (GTO) implementation of Ref.~\cite{roh95} were
carried out. 
In the ab-initio GW calculation, the wave functions have been represented by 
s and p Gaussian orbitals on carbon (decay constants: 0.12, 0.4, 1.0, and 2.8 
atomic units, i.e., 16 orbitals per atom) and by s and p orbitals on hydrogen 
(decay constants 0.15, 0.4  and 1.0, i.e., 12 orbitals per atom). The two-point 
functions occurring in the GW scheme are represented by s, p, d, and s* 
orbitals on carbon (decay constants 0.2, 0.6, 1.6, and 4.0, i.e., 40 orbitals 
per atom) and on hydrogen, as well (decay constants 0.25 and 0.7, i.e., 20 
orbitals per atom).
In both the DFTB and first principles calculations the LDA
exchange-correlation functional was used.

The results for benzene and anthracene are listed in Tab. \ref{tab_vgl} for
a small number of states around the frontier orbitals, according to the energy
partioning of Eq.~(\ref{qpe}). Focusing first on the orbital
energies, we find a very good agreement between the DFTB and the first
principles results. This might be surprising considering the limited basis set
employed in the former approach. However,  
the DFTB basis consists of optimized atomic orbitals rather than simple Gaussian
type orbitals. Moreover, the approximations underlying the DFTB method seem to
be justified due to a stable error cancellation for the $\pi$-orbitals. This
hold true to a lesser extent for $\sigma$-orbitals, like the $E_{2g}$ state in
benzene, where an error up to 0.6 eV is found. Not unexpected, difficulties
are also observed for unbound virtual orbitals like the $B_{2g}$ state in
benzene, since the description of the continuum is very sensitive to the
quality of a finite basis set.

Next, we turn to the exchange-correlation energy per orbital. Here we
find in comparison with the first principles results, that the DFTB values are in general
too positive by roughly 10 \%. However, the error even reaches 20 \% for the
$\sigma$-orbitals of benzene. We attribute this failure  to the neglect of
crystal field terms in our approach, which is 
likely to have different effects on orbitals of $\sigma$ and $\pi$
symmetry. In fact, we calculated elements of the type
$\exv{\phi_\mu^A}{v_{xc}(\rho_B)}{\phi^A_\nu}$ and found, that integrals where A
represents a hydrogen atom and B a carbon atom are significantly larger than
in the reversed situation or integrals where both A and B stand for
carbon atoms. As the latter two types of integrals occur in the calculation of
$\pi$-orbitals of the polyacenes, while the first one is important for
$\sigma$-orbitals, the missing of the crystal field terms is likely to be the
source of error here.

Considering now the exchange contribution to the self-energy $\Sigma_x^i$,
similar trends are found. Compared to the ab initio results, the DFTB values
are slightly too positive. Since the terms $\Sigma_x^i$ and $v_{xc}^i$
contribute in Eq.~(\ref{qpe}) with opposite signs to the quasiparticle
energies, a stable error cancellation is expected. A larger deviation is found
again for the $E_{2g}$ state in benzene, where an error up to 4 eV
occurs. Since the exchange integrals depend strongly on the atomic repulsion
integrals $U_{ee}$ in our approximation, the error could be reduced by
enlarging this parameter for hydrogen without loosing the good performance for
the $\pi$-orbitals. However, we hesitate to treat the $U_{ee}$ values as
empirical parameters, because of loss of transferability. Instead, one should
look for a better approximation of the two-electron integrals. In
our approximation the density $\phi_\mu(\ro)\phi_\nu(\ro)$ is represented by a
superposition of spherical charge densities. Consequently, the two-electron
integrals are given by simple monopole-monopole interactions, thus neglecting
any angular momentum dependence. A natural next step would be to include
higher order terms in a multipole expansion of the density
$\phi_\mu(\ro)\phi_\nu(\ro)$, as it is done in the semiempirical MNDO method
developed by Dewar and Thiel \cite{dew77}.

Next, the final quasiparticle energies are discussed. For the $\pi$-orbitals
the mean deviation of the DFTB results from the ab initio values is 0.4
eV, with errors decreasing when the system size is increased. As could be
already expected from the forgoing discussion, the description of
$\sigma$-orbitals is less satisfactory in the current state of
approximations. For the $E_{2g}$ orbital of benzene an error of 1.25 eV is
observed. For the unoccupied levels however, a very nice agreement between
first principles and DFTB results is evident. Clearly, this is a consequence
of an error cancellation between all terms in Eq.~(\ref{qpe}), since e.g. the
correlation contribution $\Sigma_c^i$ is systematically underestimated in the
DFTB scheme.

In this context it is interesting to investigate if a more advanced treatment
of the Dyson equation (\ref{dyson}) leads to better results. In fact, it has
been found that the associated wavefunctions of orbitals which are bound at
the DFT level of theory, but unbound at the QP level, differ
considerably. This is in contrast to the assumptions made in the derivation of
Eq.~(\ref{qpe}) from Eq.~(\ref{dyson}) and hence the full QP Hamiltonian needs
to be diagonalized in these cases and self-consistency with respect to the
energy dependence of $\Sigma$ must be achieved. Following this approach
earlier investigations of this point reported shifts of the LUMO level up to
0.8 eV \cite{roh002,ish01}. We also performed such calculations for benzene
and found that even for the $E_{2u}$ state the diagonalization changes the QP
spectrum by less than 0.01 eV. This can be understood as as consequence of the
minimal basis set we employ, which does not provide enough flexibility to
describe the relaxation towards delocalized states. Considering the energy
dependence of the self-energy, it can be stated that the approximate treatment
of Eq.~(\ref{qpe}) using the renormalization constant $Z$ is quite successful,
as we find deviations less than 0.2 eV from the self-consistent solution of
the Dyson equation.
\subsection{Size dependence of the quasiparticle gap}
\begin{figure} 
  \centering
\rotatebox{0}{ \includegraphics[scale=0.7]{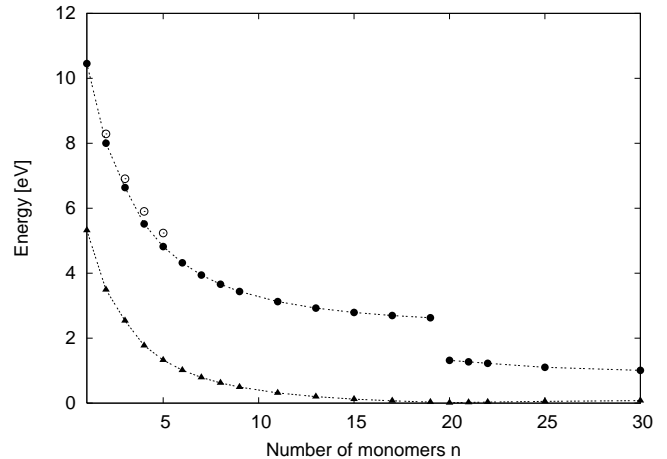}}
\caption{Quasiparticle ($\bullet$) and DFT gap
 ($\blacktriangle$) for the lowest energy conformer of the polyacenes as obtained in the
 DFTB approximation. Lines are guides to the eyes. Shown is also the
 difference of experimentally determined electron affinities and vertical
 ionization potentials ($\circ$) from Ref.~\cite{nist}, where they were
 available.\label{qpgap}}
\end{figure}
\begin{figure} 
  \centering
 \includegraphics[scale=0.3]{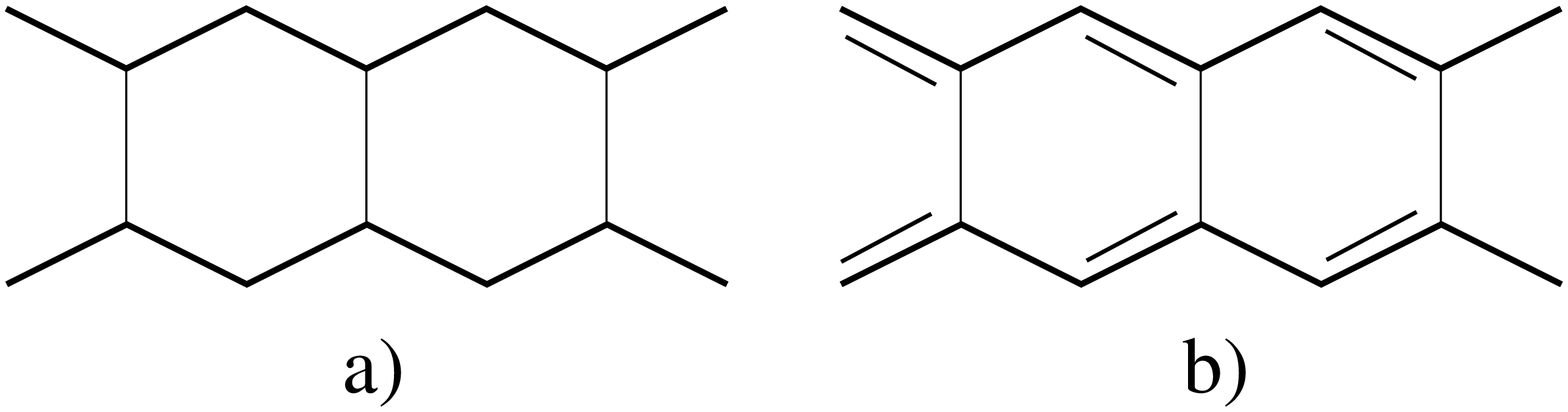}
\caption{Schematic representation of the lowest energy conformations of the
 polyacenes found in this study. The aromatic structure a) is the most stable
 for n $\le$ 19, while the Peierls (Z)-distorted structure b) is energetically
 favoured for n $\ge$ 20.\label{peierls}}
\end{figure}

After validation of the method, we now turn to a first application and analyze
in the following the quasiparticle gap $\epsilon^{QP}_{\text{gap}}=
\epsilon^{QP}_{\text{LUMO}}-\epsilon^{QP}_{\text{HOMO}}$ as a
function of chain length. The first observation which can be drawn from
Fig.~\ref{qpgap} is that the DFTB quasiparticle gap is in very nice agreement with
the experimental data, which
provides some confidence that the general trends we are looking for are
correctly described. Furthermore, Fig.~\ref{qpgap} shows that the DFT gap is
continously decreasing and almost vanishes for n = 19 monomers. As the length
increases, the geometry of the innermost part of the chain resembles more and
more that of two coupled polyacetylene chains with equal bond lengths, as
schematically depicted in Fig.~\ref{peierls}.  
The vanishing of the DFT gap can thus be understood in terms of a simple
particle-in-a-box model.

\begin{figure} 
  \centering
\rotatebox{0}{ \includegraphics[scale=0.7]{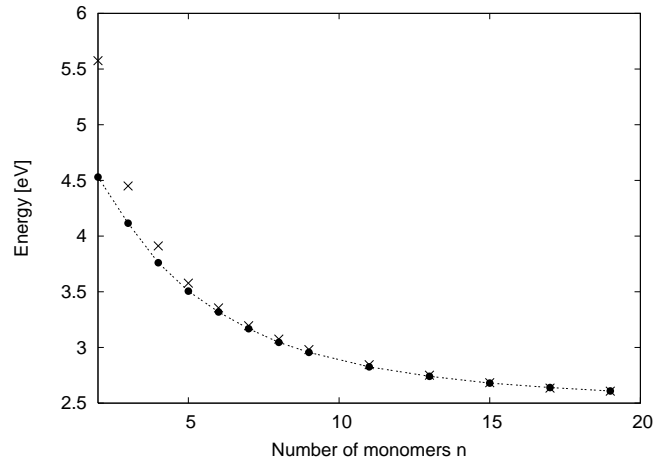}}
\caption{Fit of Eq.~(\ref{pol}) ($\times$) to the difference between QP and
  DFT gap as obtained in this work ($\bullet$).\label{qppol} }
\end{figure}

 In stark contrast to the DFT gap,
$\epsilon^{QP}_{\text{gap}}$ remains finite for increasing chain length and it
seems worthwhile to explore the physical origin of this different behaviour in
a short digression. In fact, QP energies can be directly compared to results
from photoemission and inverse photoemission measurements, i.e., they include
the effect of an extra particle, while DFT is a pure N-electron
theory. Delerue et al.~pointed out that for nanocrystals the size dependence
of the difference
$\Delta=\epsilon^{QP}_{\text{gap}}-\epsilon^{DFT}_{\text{gap}}$ can be
estimated on the basis of classical electrostatic arguments
\cite{del00}. Considering the interaction between the extra particle and its
induced surface charge on the nanocrystal, they arrived at the following
formula:
\begin{equation}
  \label{epol}
  \Delta \approx \left( 1 - \frac{1}{\epsilon(R)} \right) \frac{e^2}{R} + 0.94
  \frac{e^2}{\epsilon(R) R} \left(\frac{\epsilon(R)-1}{\epsilon(R)+1}\right) + \Delta_b,
\end{equation}
 where $\epsilon(R)$ is an effective dielectric constant and $\Delta_b$ is the
 bulk value of $\Delta$. In order to apply Eq.~(\ref{pol}) to the
 polyacenes, we took $R$ to be half of the chain length and obtained
 $\epsilon(R)$ by averaging the microscopic dielectric function in
 Eq.~(\ref{eps}). The obtained values increase from 1.72 for n = 1 to 2.14 for
 n = 19, which reflects the decreasing band gap. A fit of Eq.~\ref{pol} to our
 QP results is shown in Fig.~\ref{qppol} and leads to a value of 2.18
 eV for $\Delta_b$. Taking into account that the DFT gap is vanishing for
 $n\to\infty$, we therefore predict a QP gap of the same value for an infinite
 chain in the aromatic structure of Fig.~\ref{peierls}. Inspection of
 Fig.~\ref{qppol} further reveals that for n $>$ 4 the agreement between
 Delerue's formula and the QP results is excellent. The fact that
 Eq.~(\ref{pol}), which was developed in the context of nanocrystals also
 holds for a quasi one-dimensional system like the polyacenes is quite
 remarkable.

We now continue the discussion of Fig.~\ref{qpgap}. Between n = 19 and n = 20
HOMO and LUMO cross, which has important implications for the geometrical as
well as electronic structure of the system. Remaining in the picture of
polyacene as coupled polyacetylene, we observe that the equal C-C bond lengths
found for n $<$ 20 turn into alternating single and double bonds for larger n
as depicted in Fig.~\ref{peierls}, i.e., the system undergoes a Peierls
distortion. In contrast to polyacetylene, where the bond alternation is found
to be around 0.08 $\AA$ \cite{suh83}, the effect is much weaker here with a
value of less than 0.008 $\AA$. Nevertheless, the dimerization leads like in
polyacetylene to an opening of the DFT gap, which tends towards a small but
finite value for the infinite chain. Inspection of Fig.~\ref{qpgap} shows that
also the quasiparticle gap differs significantly for the distorted and
undistorted structure. This fact could be used in experiment to discriminate
between both polymorphs, since we find in line with the MP2 results of
Cioslowski \cite{cio93}, that the two forms are energetically quite close and
should therefore coexist in real samples. We should mention however, that up
to now only polyacenes up to n = 6 could be isolated, since larger chains are
highly vulnerable to photooxidation \cite{not98}. 

The results of this section can be summarized as follows. First, the aromatic
from of the infinite polyacene is predicted to be metallic at the DFT level of
theory, but semiconducting at the GW level. It thus provides another example
besides bulk Ge \cite{hyb86}, where the DFT gap is not only quantitatively but
also qualitatively wrong. Second, the difference between the DFT and QP gap
can be understood in terms of the interaction between an extra particle -
which is missing in DFT - and the charge it induces on the molecular surface.
Third, the Peierls distorted polyacene conformer is energetically favoured
only for very long chains and posesses a QP spectrum which is markedly
different from the aromatic form. This also underlines the usefulness of
approximate GW schemes, since in a first principles context the Peierls
transition found here might not be noticed due to the limited treatable system
size \cite{rem4}.

\section{Numerical considerations}
In the following, we shortly discuss the numerical efficiency of our
approximations. The method scales like $N^2N_l^2$, where N is the number of
basis functions and $N_l$ is the number of spherically averaged
basisfunctions, that are used in the representation of two-point functions
($N_l < N$). This has to be compared to first principles implementations,
which usually scale like $N^4$. An additional reduction of computation time is
obtained, since we use a minimal basis of optimized atomic orbitals, while in
a first principles framework a larger number of primitive orbitals is
required. Moreover, the scaling prefactor is reduced in the DFTB scheme,
because the necessary integrals are either precalculated and tabulated or
approximated by simple functions. As an example, the first principles
evaluation of the QP spectrum of anthracene took 170 minutes on a Pentium Xeon
2.20 GHz processor (including 120 minutes for the DFT part of the
calculation), compared to less than 1 second on a Pentium 4 with 2.40 GHz in
our approach. The largest structure we studied, the n = 30 polyacene with 186
atoms took 10 minutes. The limiting factor in the calculation of very large
systems is therefore not the computation time but rather the memory
requirement. We try to circumvent this problem by computing memory intensive
quantities like the overlap charges $q^{ij}_\mu$ on-the-fly in a direct way.
 
\section{Concluding remarks}
In this work we presented a method to perform quasiparticle calculations of
molecular systems at a highly reduced computational cost compared to first
principles implementations. The scheme was applied to hydrocarbons,
but it can be easily extended to other elements, since all required
parameters are calculated from first principles. The various
approximations of the method are intended to be as
consistent as possible with the underlying DFTB approach to allow for a stable
error cancellation. For benzene and anthracene the results are indeed
comparable with higher level calculations with the exception of
$\sigma$-orbitals. Here, ways to overcome the deficiencies were
outlined. Nevertheless, we think that the scheme could be useful already at
the present stage, since e.g. for optical spectra or in the electronic
transport only a few states around the Fermi level are active and dominate the
physical properties of a system.

\section*{Acknowledgements}

The authors would like to thank Alessandro Pecchia and Alessio
Gagliardi for fruitful discussions related to this work. Further, the EC-Diode-Network is gratefully
acknowledged for financial support and T.A.N is much obliged for using the
computer facilities  at the German Cancer
Research Center in Heidelberg.


\begin{thebibliography}{99}
\bibitem{rem1} Please note that in the original formulation of
   Ref.~\cite{els98} the Mulliken charges were not resolved according to
   the angular momentum. The higher flexibility offered by the formulation in
   this work has only marginal influence on ground state properties, but is
   crucial for the description of exchange integrals that are needed here.
\bibitem{rem3} Traditionally, the $U^H$ parameters are calculated from an
   umcompressed wavefunction in the DFTB approach. To be consistent, we obtain
   the $U^{ee}$ parameters in the same way here, although this is somewhat
   inconsistent with the treatment of the exchange energy, where values from a
   compressed wavefunction are used.
\end{thebibliography}
\end{document}